\documentclass[conference]{IEEEtran}

\usepackage{amsmath,amssymb}
\usepackage{array}
\usepackage{booktabs}
\usepackage[style=ieee,maxnames=4,minnames=3,maxbibnames=3]{biblatex}

\usepackage{graphicx}
\usepackage{xcolor}
\usepackage{bm}

\begin{document}

\title{
JEPA-Inspired Latent Field Prediction for Sparse-Pilot MIMO-OFDM Channel Reconstruction}
\author{
\IEEEauthorblockN{
Yuzhi~Yang\IEEEauthorrefmark{1}\IEEEauthorrefmark{2},
Brahim~Mefgouda\IEEEauthorrefmark{1},
Hang~Zou\IEEEauthorrefmark{1},
Lina~Bariah\IEEEauthorrefmark{1},~
Anis~Bara\IEEEauthorrefmark{1},
}\IEEEauthorblockN{
Yuhuan~Lu,\IEEEauthorrefmark{3},
Hao~Zhang\IEEEauthorrefmark{1},~
and M\'erouane~Debbah\IEEEauthorrefmark{1}
}
\IEEEauthorblockA{
\IEEEauthorrefmark{1}Research Institute for Digital Future, Khalifa University, 127788 Abu Dhabi, UAE
}
\IEEEauthorblockA{
\IEEEauthorrefmark{2}School of Information Science, Japan Advanced Institute of Science and Technology, Nomi, Ishikawa, Japan}
\IEEEauthorblockA{
\IEEEauthorrefmark{3}the Faculty of Applied Sciences, Macao Polytechnic University, Macao SAR, China}
\IEEEauthorblockA{
E-mail: {yangyzis@jaist.ac.jp}
}
\vspace{-0.8cm}
}

\maketitle

\begin{abstract}
Accurate acquisition of full channel state information (CSI) is essential for multiple-input multiple-output orthogonal frequency-division multiplexing (MIMO-OFDM) wireless systems, while dense pilot-based estimation incurs substantial overhead. Directly minimizing the prediction error over raw CSI coefficients is fragile in realistic propagation environments, since small-scale fading, mobility, local scattering, and small spatial perturbations can induce large phase-sensitive variations. Nevertheless, wireless channels contain strong spatial, temporal, and frequency-domain correlations that can be more predictably represented in a latent propagation-field domain. Motivated by this observation, we propose a joint-embedding predictive architecture (JEPA)-inspired field world model, defined as a latent predictive model of the wireless propagation field, for sparse-pilot channel reconstruction. The model maps CSI observations into latent time-frequency tokens, predicts masked or future latent field states from historical multi-band context, and fuses the predicted latent prior with sparse noisy current pilots to reconstruct the full MIMO-OFDM channel. Ray-tracing experiments show that the proposed method improves normalized mean square error (NMSE) over pilot-only and direct reconstruction baselines, supporting latent field prediction as an effective prior for sparse-pilot channel acquisition.
\end{abstract}

\section{Introduction}
In multiple-input multiple-output orthogonal frequency-division multiplexing (MIMO-OFDM) wireless systems, accurate channel state information (CSI) is essential for coherent detection, beamforming, precoding, and resource allocation. In practical systems, CSI is usually acquired with pilot signals inserted over the time-frequency resource grid. However, when the channel varies across time slots and frequency bands, repeatedly transmitted pilots may consume considerable spectral resources. This motivates pilot-efficient channel acquisition methods that exploit the correlation among channel observations to reconstruct the full CSI from sparse current pilots. 

However, direct channel prediction is fragile in realistic wireless environments. The observed channel response contains both relatively stable propagation structures and rapidly varying small-scale fading components, while mobility, local scattering, and small spatial perturbations can induce large phase-sensitive variations. As a result, it becomes fragile to generate satisfactory raw CSI coefficients with simple prediction methods. Nevertheless, the underlying propagation field contains strong spatial, temporal, and frequency-domain correlations, which introduce an inherent strong correlation between resource blocks. This suggests that prediction should be performed in a suitable latent propagation-field domain and used as a structured prior, rather than treated as a complete replacement for pilot-based channel estimation.

Time- and frequency-domain channel prediction has been widely studied as an effective way to acquire future channel state information with reduced signaling overhead. Classical model-based methods can exploit temporal correlation, whereas their performance strongly depends on the accuracy of the assumed channel dynamics. Meanwhile, recent learning-based methods learn the mapping from historical CSI to future CSI directly from the data, which usually outperforms classical methods. As summarized in \cite{liu2024llm4cpadaptinglargelanguage}, recurrent networks, transformer-based predictors, large language model-based predictors, and wireless foundation models have been explored to capture richer space-time-frequency correlations of wireless channels. These studies demonstrate the potential of deep learning for channel prediction. However, many prediction-oriented methods still operate directly on raw CSI coefficients, without explicitly separating predictable propagation-field structure from phase-sensitive instantaneous fluctuations or using predicted latent states as priors for sparse-pilot full-channel reconstruction.

Meanwhile, world models have recently emerged as a promising paradigm for learning compact representations of complex dynamical systems \cite{vjepa}. In particular, their success in video prediction and reconstruction shows that high-dimensional observations often contain latent physical structures that can be predicted more effectively in an abstract representation space than directly in the raw observation space \cite{vjepa,xiang2025pan}. Instead of capturing pixel-level details, a world model aims to capture the underlying state and its temporal evolution, enabling prediction, completion, and reconstruction from partial observations. This perspective is naturally relevant to wireless communications. Although instantaneous small-scale fading may fluctuate rapidly, channel evolution is still governed by hidden environmental dynamics. Therefore, a world model can be used to learn a latent channel representation that emphasizes the predictable large-scale relationship across time and frequency, while leaving fine-grained signal details to be recovered with the aid of sparse raw observations \cite{twm}. Recent wireless JEPA studies \cite{naoumi2026structuredlatentdynamicswireless} have explored latent dynamics for channel representation and charting, whereas sparse-pilot full-channel reconstruction with predicted latent priors remains less explored.

In this work, we further adopt dense predictive supervision and deep self-supervision over multiple intermediate vision transformer (ViT) layers \cite{vjepa21}. Dense predictive supervision constrains the predicted target field at the token level rather than only through a global latent vector, which is important for preserving local subcarrier-antenna and time-frequency structure. Intermediate-layer self-supervision encourages the model to learn propagation-field representations at multiple abstraction levels, improving the stability of latent prediction. Therefore, the proposed field world model predicts structured latent propagation states rather than raw CSI coefficients, and the predicted latent field is later used as a prior for sparse-pilot channel reconstruction.

Motivated by the above observation, we propose a joint-embedding predictive architecture (JEPA)-inspired field world model (FWM) for sparse-pilot MIMO-OFDM channel reconstruction. Here, FWM denotes a latent predictive model of the wireless propagation field. Unlike methods that directly extrapolate raw CSI coefficients, the proposed framework maps CSI observations into latent time-frequency tokens and predicts masked or future latent field states from historical multi-band context. The predicted latent field is not required to preserve all instantaneous CSI details. Instead, it is designed to capture stable and predictable propagation structure. For downstream reconstruction, this predicted latent state is used as a structured prior and fused with sparse noisy current pilots to recover the full MIMO-OFDM channel.

The main contributions of this paper are summarized as follows:
\begin{itemize}
    \item We formulate sparse-pilot MIMO-OFDM channel reconstruction as a latent-field-aided reconstruction problem, where predictable propagation-field states are learned from historical multi-band time-frequency CSI and used as priors for current channel recovery.

    \item We develop a JEPA-inspired field world model that predicts masked or future channel states in a compact latent propagation-field domain, avoiding direct coefficient-level extrapolation of phase-sensitive raw CSI.

    \item We design a pilot-assisted reconstruction framework that fuses the predicted latent field prior with sparse noisy current pilots to recover the full channel. The numerical results also show that overly detail-preserving auxiliary objectives can reduce downstream reconstruction usefulness.
\end{itemize}
\vspace{-0.3cm}
\section{Proposed Method}
\begin{figure*}[t]
    \centering
    \includegraphics[width=0.7\linewidth]{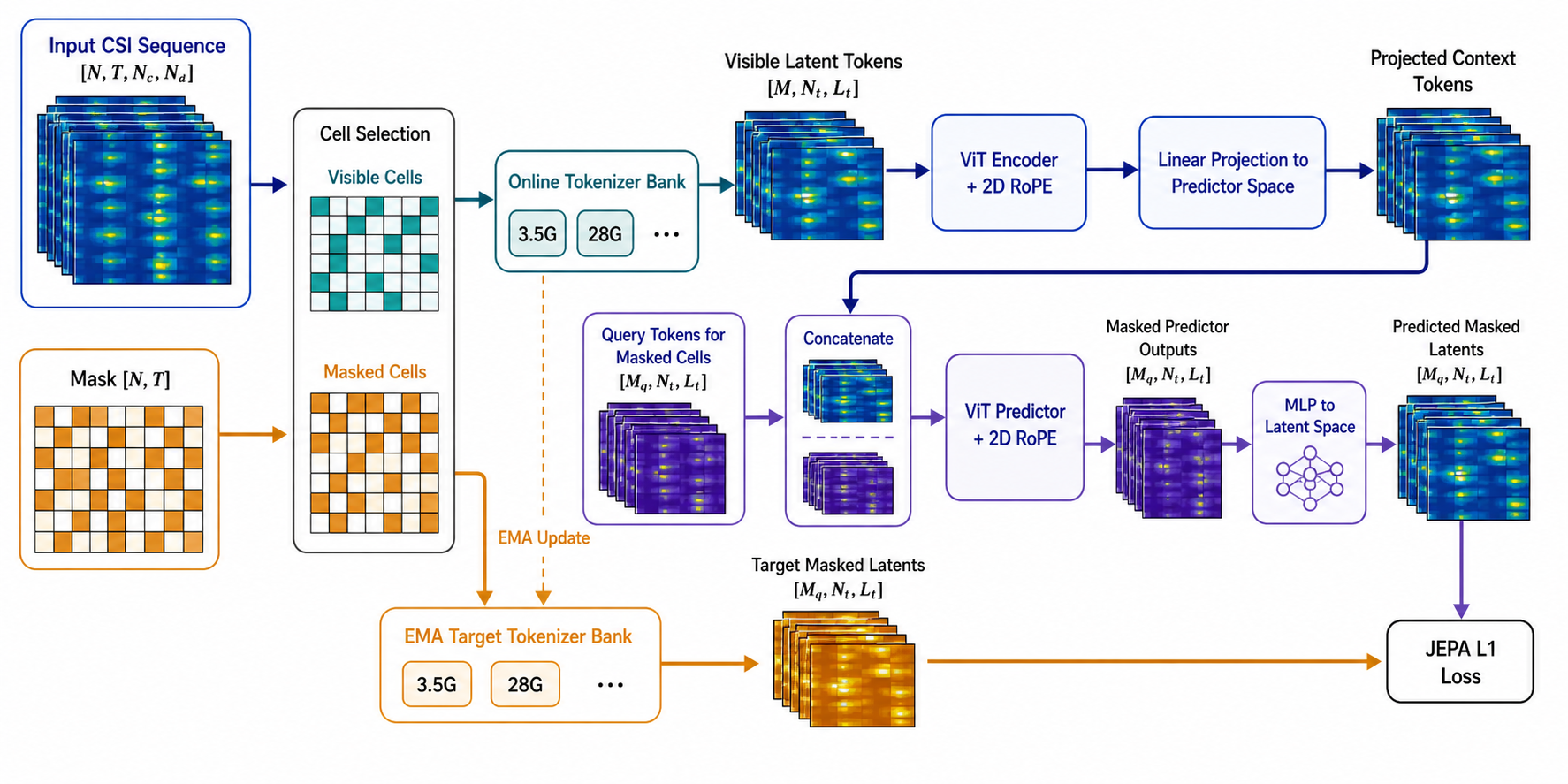}\vspace{-0.5cm}
    \caption{Overview of the time-subband prediction JEPA pipeline.}
    \label{fig:timeband_jepa_pipeline}\vspace{-0.5cm}
\end{figure*}

\subsection{System Overview}

In this paper, we propose a JEPA-based FWM for time-frequency channel prediction and pilot-assisted channel reconstruction. The key idea is to treat each subband-time CSI observation as a cell in a time-frequency propagation field. Instead of directly extrapolating raw CSI coefficients, the proposed field world model maps each CSI cell into a latent token representation and performs prediction in the latent propagation-field domain. The predicted latent state is then used as a structured prior for downstream reconstruction from sparse current pilots.

The overall architecture consists of three modules. First, a CSI tokenizer converts each CSI cell into a group of latent tokens. Second, a JEPA-style two-stage ViT backbone predicts the latent representations of masked or future cells from visible historical cells. Finally, for downstream channel estimation, the predicted latent prior is combined with sparse pilot observations and fed into a downstream head to reconstruct the full CSI.

\subsection{JEPA-Based Time-Frequency Latent Prediction}
The JEPA pretraining pipeline is generally depicted in Fig. \ref{fig:timeband_jepa_pipeline}, which considers a typical channel latent prediction problem in the time and frequency fields. Let $H_{n,t}\in\mathbb{C}^{N_{\mathrm{c}}\times N_{\mathrm{a}}}$ denote the complex CSI matrix at subband $n$ and time slot $t$, where $N_{\mathrm{c}}$ is the number of subcarriers and $N_{\mathrm{a}}$ is the number of antenna elements. The full channel trajectory is represented as $\mathcal{H}\in\mathbb{C}^{T\times N_\mathrm{b}\times N_{\mathrm{c}}\times N_{\mathrm{a}}}$, where $T$ and $N_\mathrm{b}$ denote the numbers of time slots and subbands, respectively. We define $\mathcal{A}$ as the set of all time-subband cells, $\mathcal{R}$ as the set of visible cells, and $\mathcal{M}=\mathcal{A}\setminus\mathcal{R}$ as the set of masked cells to be predicted. Each visible CSI cell is mapped into $K$ latent tokens of dimension $D$ by a tokenizer
\begin{equation}
    L_{n,t}
    =
    f_{\mathrm{tok}}(H_{n,t};\theta)
    \in \mathbb{R}^{K\times D},
    \quad (n,t)\in\mathcal{R}.\label{eq1}
\end{equation}
The latent tokens are organized as a time-frequency field. We encode the global time-subband coordinates using two-dimensional positional encoding, while the local index of each token within a cell is represented by a learnable embedding. In each training sample, all cells at the final time slot $T$ are masked to simulate the next-step channel-field prediction. Historical cells are further randomly masked to encourage completion from incomplete time-frequency context, corresponding to the case where not all previous resource blocks are available for the target user.

Following the JEPA-style prediction paradigm \cite{vjepa21}, the backbone contains a context encoder and a latent predictor. The context encoder processes only the visible cells:
\begin{equation}
    \mathcal{Z}_{\mathcal{R}}
    =
    f_{\mathrm{enc}}
    \left(
        \{L_{n,t}\}_{(n,t)\in\mathcal{R}},
        \{(n,t)\}_{(n,t)\in\mathcal{R}};
        \phi_{\mathrm{enc}}
    \right),
\end{equation}
where $\mathcal{Z}_{\mathcal{R}}$ denotes the contextualized latent embeddings of the observed time-frequency cells. The masked cells are not provided to the encoder.

For each masked cell $(n,t)\in\mathcal{M}$, we introduce a learnable query token group $Q_{n,t}\in\mathbb{R}^{K\times D_p}$. The predictor takes the projected context embeddings and all query tokens as input:
\begin{equation}
\begin{aligned}
    Y
    =
    f_{\mathrm{pred}}
\big(
        \big[
        g_{\mathrm{in}}(\mathcal{Z}_{\mathcal{R}}),&
        \{Q_{n,t}\}_{(n,t)\in\mathcal{M}}
        \big],\\&\left.
        \{(n,t)\}_{(n,t)\in\mathcal{R}\cup\mathcal{M}};
        \phi_{\mathrm{pred}}
    \right).
\end{aligned}
\end{equation}
The output tokens corresponding to the masked cells are then mapped back to the tokenizer latent space through a linear layer:
\begin{equation}
    \hat{L}_{n,t}
    =
    f_{\mathrm{out}}(Y_{n,t})
    \in \mathbb{R}^{K\times D},
    \quad (n,t)\in\mathcal{M}.\label{eq4}
\end{equation}
In this way, the model learns to infer the latent state of missing or future CSI cells from the observed time-frequency context, rather than extrapolating raw channel coefficients directly.

\subsection{Self-Supervised Training Objective}

The model is trained by masked latent prediction. To provide stable targets, we use a momentum tokenizer with parameters $\bar{\theta}$, which has the same architecture as the online tokenizer. For each masked cell, the target latent is
\begin{equation}
    L_{n,t}^{\mathrm{tar}}
    =
    f_{\mathrm{tok}}(H_{n,t};\bar{\theta}),
    \quad (n,t)\in\mathcal{M}.
\end{equation}
The momentum tokenizer is updated by exponential moving average:
\begin{equation}
    \bar{\theta}
    \leftarrow
    \mu\bar{\theta}
    +(1-\mu)\theta,
\end{equation}
where $\mu$ is the momentum coefficient. The main JEPA loss is defined as
\begin{equation}
    \ell_{\mathrm{JEPA}}
    =
    \frac{1}{|\mathcal{M}|}
    \sum_{(n,t)\in\mathcal{M}}
    \frac{1}{KD}
    \left\|
        \hat{L}_{n,t}
        -
        L_{n,t}^{\mathrm{tar}}
    \right\|_1 .
\end{equation}

\begin{figure}[t]
    \centering
    \includegraphics[width=0.85\linewidth]{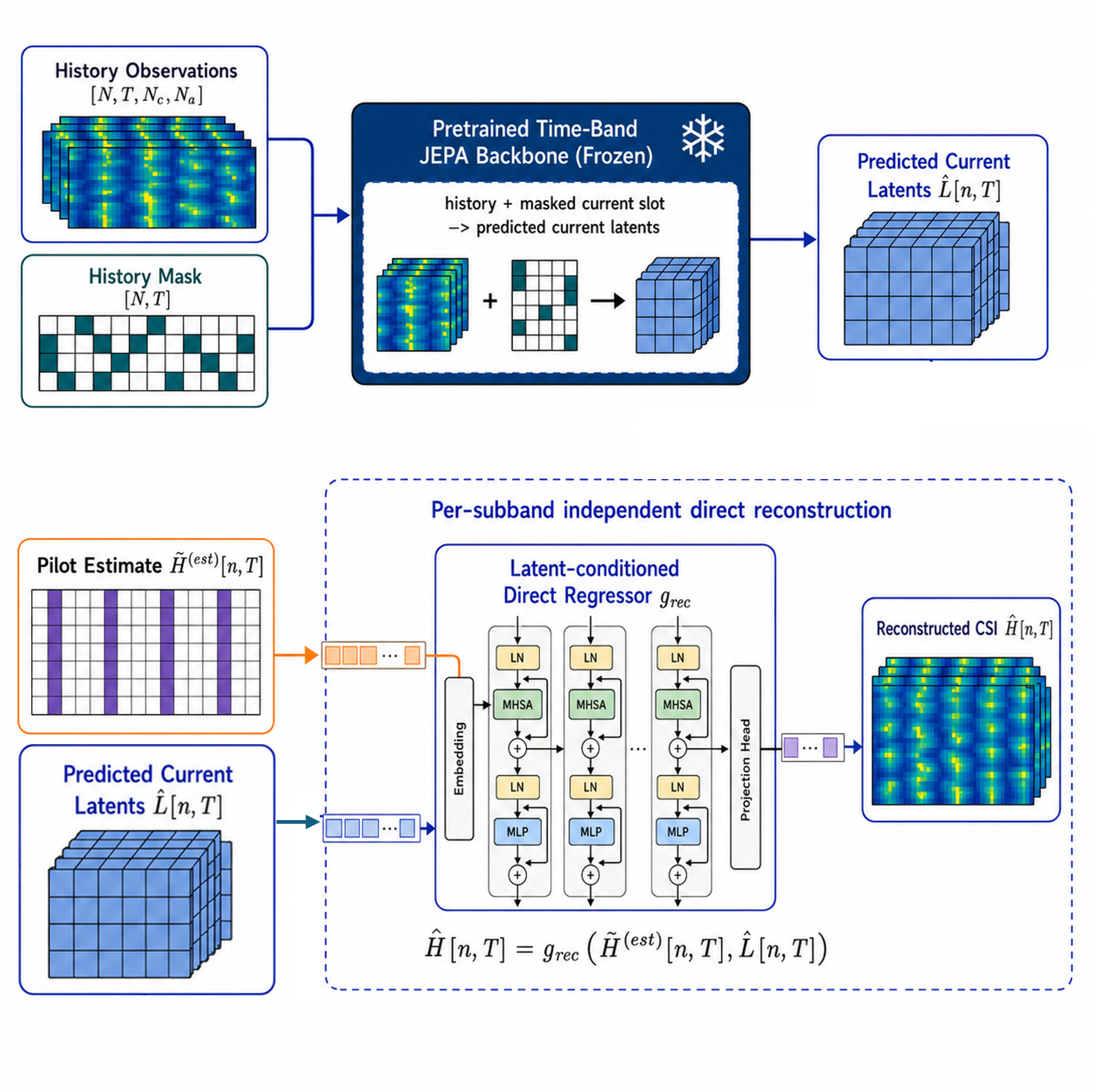}\vspace{-0.5cm}
    \caption{Pilot-aware channel prediction and reconstruction structure.}
\label{fig:pilot_aware_channel_prediction}\vspace{-0.5cm}
\end{figure}
A pure latent prediction objective may lead to representation collapse, where all CSI observations are mapped to nearly constant latent codes \cite{naoumi2026structuredlatentdynamicswireless}. To avoid this, we introduce a regularization to encourage statistical diversity of the latent field. Specifically, we use a covariance loss to reduce redundancy among latent dimensions
\begin{equation}
    \ell_{\mathrm{cov}}
    =
    \left\|
        C-\mathrm{diag}(C)
    \right\|^2_F ,
\end{equation}
where $C$ is the covariance matrix of the flattened latent features over the current batch.

In addition to the main JEPA-covariance objective, we evaluate several CSI-domain grounding objectives as ablations, including direct CSI recovery, amplitude recovery, and conditional diffusion grounding. These ablations are used to examine whether forcing the latent representation to preserve more CSI-domain details improves or degrades downstream reconstruction. Since diffusion grounding predicts injected noise from a corrupted CSI sample, it imposes a weaker deterministic recovery constraint on the latent code than direct CSI recovery. Among the considered grounding objectives, direct CSI recovery imposes the strongest detail-preservation constraint, while diffusion grounding is the weakest. These objectives share the generic form
\begin{equation}
    \ell_{\mathrm{aux}}
    =
    \frac{1}{|\mathcal{M}|}
    \sum_{(n,t)\in\mathcal{M}}
    \left\|
        f_{\mathrm{aux}}\left(U_{n,t}^{\mathrm{aux}};\psi\right)
        -
        V_{n,t}^{\mathrm{aux}}
    \right\|_2^2 ,
\end{equation}
where \((U_{n,t}^{\mathrm{g}},V_{n,t}^{\mathrm{g}})\) specify the input and target of each grounding variant, and $\psi$ denotes the parameters in the auxiliary method which is trained alongside other trainable parameters. Direct CSI recovery uses the latent code to recover the normalized CSI, such that $U_{n,t}^{\mathrm{aux}}=L_{n,t}$ and $V_{n,t}^{\mathrm{aux}}=H_{n,t}$. The amplitude recovery replaces the target by channel amplitude, such that $U_{n,t}^{\mathrm{aux}}=L_{n,t}$ and $V_{n,t}^{\mathrm{aux}}=\mathrm{Amp}(H_{n,t})$, where $\mathrm{Amp}(H)$ denotes the element-wise amplitude (absolute value) of $H$. The diffusion grounding uses the latent code together with a noisy CSI sample and diffusion step to predict the injected noise, such that $U_{n,t}^{\mathrm{aux}}=(\bar{H}_{n,t}^{(\tau)},L_{n,t},\tau)$, $\bar{H}_{n,t}^{(\tau)}=\alpha_\tau H_{n,t}+\sqrt{1-\alpha_\tau^2}\epsilon_{n,t}$ with a random Gaussian white noise $\epsilon_{n,t}$, and $V_{n,t}^{\mathrm{aux}}=\epsilon_{n,t}$. The diffusion-based realization is similar to that in \cite{xiang2025pan}.

The overall pretraining objective is thus
\begin{equation}
    \ell_{\mathrm{pre}}
    =
    \ell_{\mathrm{JEPA}}
    +
    \lambda_{\mathrm{cov}}\ell_{\mathrm{cov}}
    +
    \lambda_{\mathrm{g}}\ell_{\mathrm{g}}.
\end{equation}
The final model uses $\ell_{\mathrm{JEPA}}+\lambda_{\mathrm{cov}}\ell_{\mathrm{cov}}$, while $\ell_{\mathrm{g}}$ is enabled only in ablation studies.

\section{Pilot-Assisted Channel Reconstruction With Latent Priors}
The JEPA backbone provides a structured latent representation of the current wireless field. However, the predicted latent field is not intended to replace current pilot observations. Instead, it provides a structured prior that captures predictable propagation information from historical time-frequency context. We therefore evaluate the learned FWM through a pilot-assisted channel reconstruction task, where the current full CSI is recovered from sparse noisy pilots containing small-scale information conditioned on the predicted latent state.

Based on the JEPA latent representation and its prediction $\hat{L}_{n,t}$ from historical channel observations $H_{n,t}\in\mathbb{C}^{N_{\mathrm{c}}\times N_{\mathrm{a}}}$ over previous time steps through \eqref{eq1}-\eqref{eq4}, we formulate a reconstruction task for the current channel $H_{n,T}$.

Meanwhile, the receiver obtains an initial rough and partial estimation of the current channel based on pilots. This estimation is sparse and noisy, and can be modeled as
  \begin{equation}
      \tilde{H}_{n,t}^{(\mathrm{est})}
      =
      P_{n,t}\odot (H_{n,t}+N_{n,t}),
  \end{equation}
where $P_{n,t}$ is a binary pilot mask, $N_{n,t}$ denotes the estimation noise, and $\odot$ denotes element-wise multiplication. We note that we use the direct tokenizer output $H_{n,t}$ corresponding to all resource blocks without passing into the JEPA predictor during training.

The final reconstruction is then produced by a latent-conditioned direct regressor that combines the predictive latent prior with the estimation input
  \begin{equation}
      \hat{H}_{n,t}
      =
      g_{\mathrm{rec}}
      \left(
          \tilde{H}_{n,t}^{(\mathrm{est})},
          \hat{L}_{n,t}
      \right).
  \end{equation}
In this way, $\hat{L}_{n,t}$ provides a latent-domain prior, while $\tilde{H}_{n,t}^{(\mathrm{est})}$ preserves instantaneous measurement evidence from the pilot-based estimation. During downstream training, the reconstruction network is trained with teacher-forced latent conditions obtained by directly tokenizing the current CSI using the frozen tokenizer. During evaluation, these teacher-forced latent conditions are replaced by the predicted latent states produced by the FWM from historical visible cells. This strategy increases the number and diversity of supervised reconstruction pairs and decouples the learning of pilot-latent fusion from latent prediction. Especially, no current full CSI is used at inference.

The downstream model is trained by direct regression with
  \begin{equation}
      \ell_{\mathrm{rec}}
      =
      \mathbb{E}
      \left[
          \left\|
              \hat{H}_{n,t}
              -
              H_{n,t}
          \right\|_2^2
      \right].
  \end{equation}
In the current implementation, the binary mask is not provided as a separate input. Since the sparse pilot observations are zero-filled according to the stride-based pilot pattern, the observed positions are implicitly encoded in the input.
The final recovery result is therefore obtained in a single forward pass from the historical latent prior and the initial channel estimation. During downstream training, the pretrained tokenizer and latent predictor are frozen, and only the reconstruction network \(g_{\mathrm{rec}}\) is optimized. The downstream network is trained with an MSE loss, while reconstruction quality is reported using NMSE.
\section{Numerical Results}
\begin{figure}[t]
    \centering
    \includegraphics[width=0.8\linewidth]{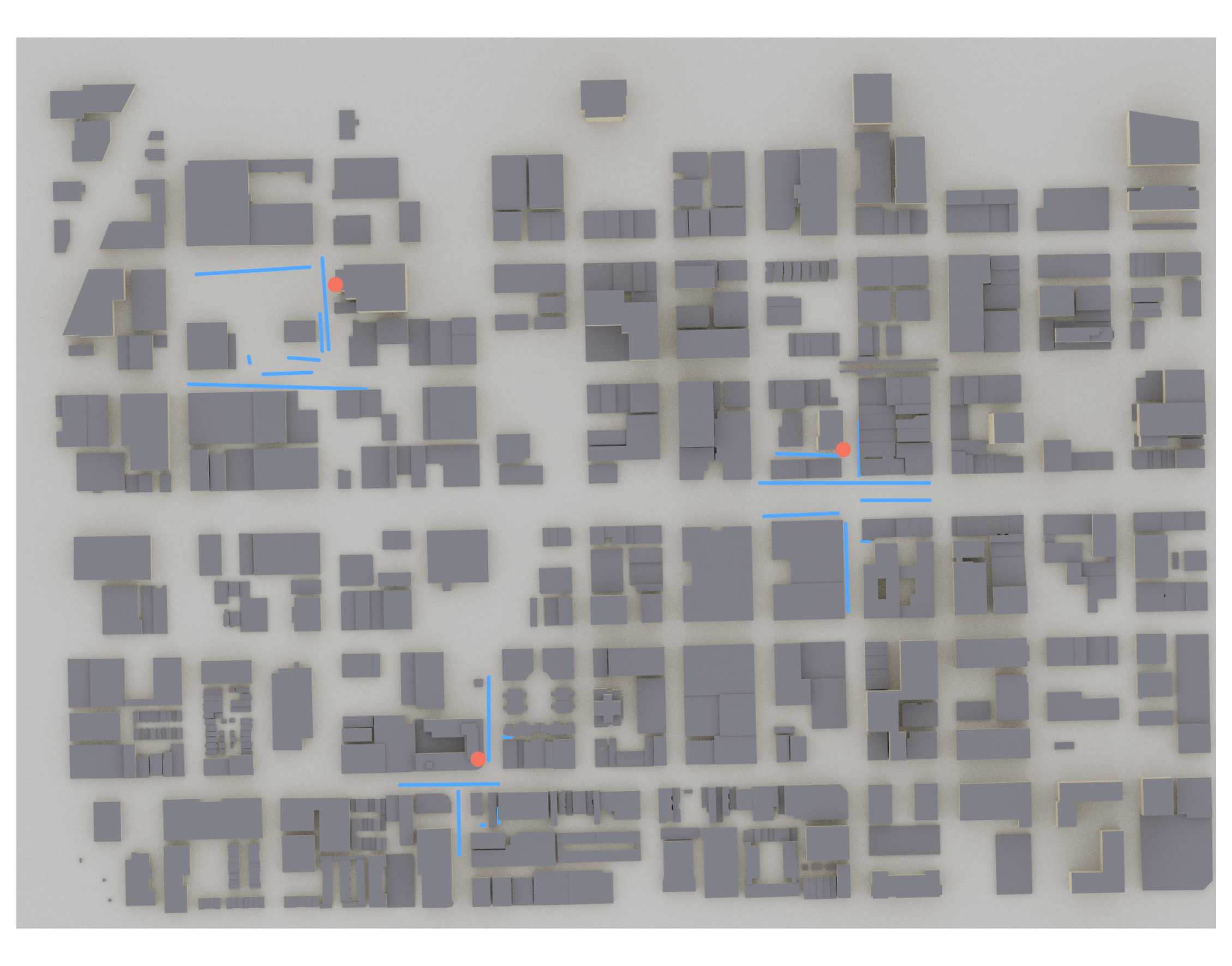}
    \vspace{-0.5cm}
    \caption{The scenario considered for generating dataset based on the actual maps in Chicago within a range of around 1 km$^2$. The red point indicates the BSs and blue lines are UE paths.}
    \label{fig:scenario}
    \vspace{-0.5cm}
\end{figure}\subsection{Dataset Generation}
\begin{table}[t]
\centering
\caption{Dataset and system configuration.}
\label{tab:dataset}
\begin{tabular}{l|l}
\hline
Item & Configuration \\
\hline
3.5 GHz subbands & 3.1, 3.3, 3.5, 3.7, 3.9 GHz \\
3.5 GHz subcarrier spacing & 480 kHz \\
28 GHz subbands & 27, 28, 29 GHz \\
28 GHz subcarrier spacing & 7.5 MHz \\
Subcarriers per subband & 32 \\
BS antennas & Two co-located $4\times4$ arrays \\
\hline
\end{tabular}\vspace{-0.3cm}
\end{table}
We generate the dataset using Sionna \cite{sionna} over the city-center scenario of Chicago. There are totally 3 BSs located on the top of the buildings. We randomize 200 trajectories for each BS. The starting and ending points are located on the ground within a range of horizontal distances of 100 m from the corresponding BS. Each trajectory contains 16 points which are almost uniformly distributed in a line. In particular, the first and last points are located with a distance within 1-10 meters, and all other points are uniformly located between them. The velocities are calculated accordingly such that the time difference between each adjacent points is 50 ms. After that, each UE position is perturbed by a Gaussian shift with standard deviation 5 cm to imitate actual moving and avoid unrealistically regular phase evolution. We consider a total of 8 different subbands, whose details are shown in Table \ref{tab:dataset}.
Therefore, each channel realization is represented as a complex tensor of dimension $16\times8\times32\times16$, corresponding to time samples, subbands, subcarriers, and antennas. We first split complete UE trajectories into disjoint training and testing sets at the trajectory level. All temporal slices are then generated within each split independently, ensuring that no trajectory contributes samples to both training and testing.
\subsection{Visibility Mask Generation}
We employ a progressive history masking strategy during JEPA training. Random masking is applied only to historical time steps, whereas the final time step is always fully masked so that the predictor must infer the target state entirely from preceding observations. To begin with, we use a fixed masking ratio at the beginning of training, as the JEPA tokenizer is not powerful enough at that time. In addition, we gradually apply a stochastic target distribution throughout the training procedure.

Specifically, for epoch $e$, the masking rate is linearly interpolated between $0.3$ and a sample drawn from a Beta distribution over $[0.1,0.9]$, namely
\begin{equation}
      p^{(e)}
      =
      (1-\lambda_e)\cdot 0.3
      +
      \lambda_e \cdot (0.1+0.8u),
      \;\; u\sim\mathrm{Beta}(4.0,1.2),
\end{equation}
where $\lambda_e$ increases linearly from 0 to 1. This design progressively increases task difficulty while preserving full masking of the final time step throughout pretraining.

\subsection{Network Topology and Training Settings}
In implementation, each complex CSI cell is represented by stacking its real and imaginary components as two channels.
For the main JEPA pretraining task, we adopt a CNN-style tokenizer head and a two-layer ViT-style predictor. Specifically, the tokenizer starts with a $3\times 3$ convolutional stem that maps the two-channel input to 32 channels. It is followed by 4 residual CNN blocks with channel transitions $32\rightarrow 64\rightarrow 128\rightarrow 128 \rightarrow 128$. A $2\times$ stride is applied to the second residual module, while others do not contain downsampling. Afterwards, each token map is then flattened and projected by a 2-layer MLP to a 192-dimensional token embedding, and a layer normalization layer is used to avoid collapsing.

The first module of the ViT-style predictor is composed of 4 transformer layers with 8 attention heads, following by a linear layer mapping the latent dimension to 256. Then, a shallow ViT-style stack with 2 transformer layers with 4 attention heads conducts the prediction, followed by a two-layer MLP network mapping them back to the 192-dimensional latent space with 2 independent tokens.

For the auxiliary regularization heads, we use the following simple modules. The masked recovery head is a CNN decoder symmetric to the tokenizer head. Meanwhile, the amplitude recovery branch uses exactly the same decoder and differs only in the training loss, which is applied to the recovered amplitude instead of the complex-valued output itself, allowing phase shifting. The diffusion regularization head is a lightweight token-conditional DiT-style network with hidden dimension 128, depth 3, and 8 attention heads, trained with a standard 1000-step noise schedule.

\begin{figure}[t]
    \centering
    \includegraphics[width=0.9\linewidth]{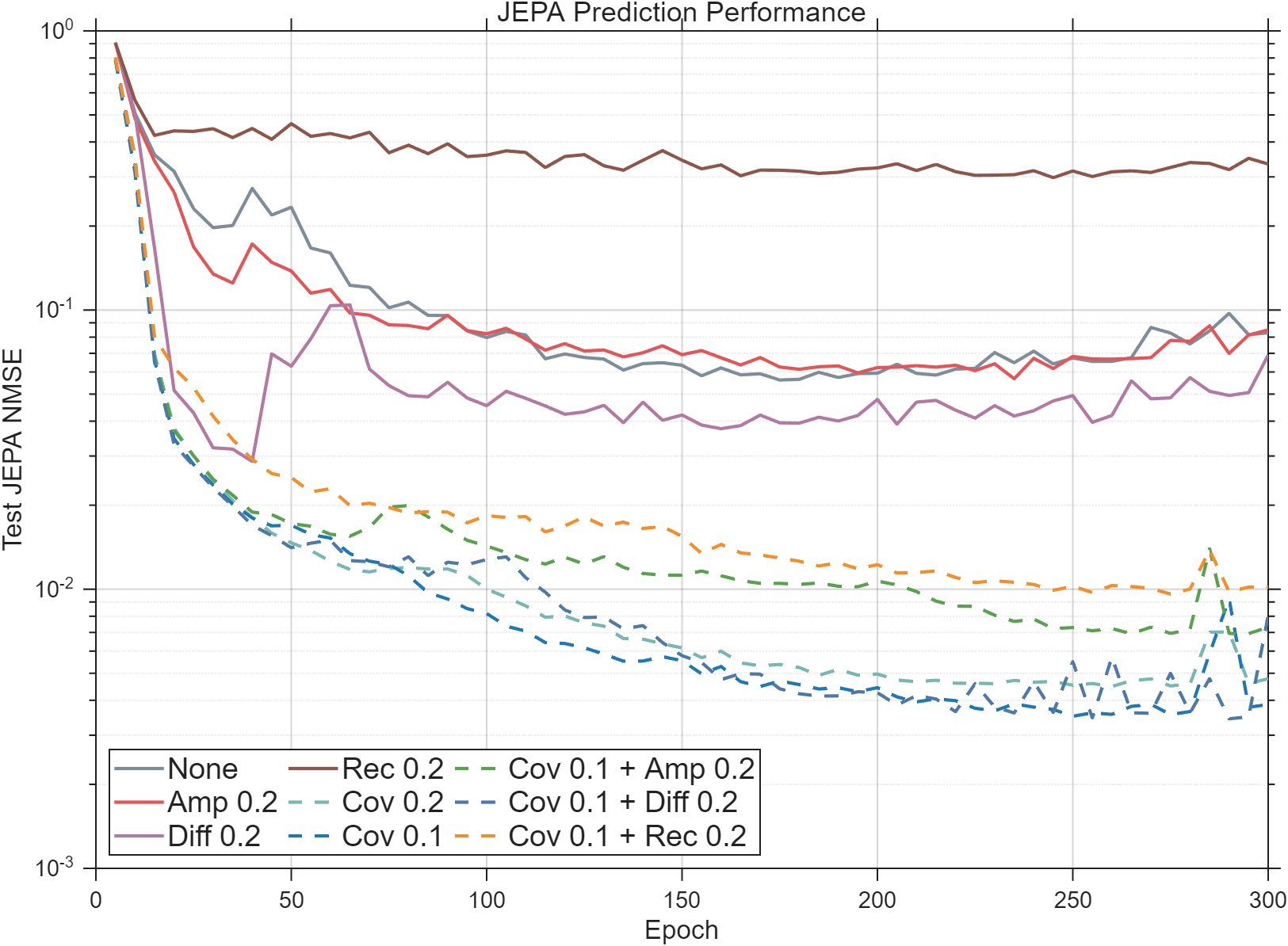}
    \caption{Pretraining results for the JEPA model.}
    \label{fig:jepa_results}\vspace{-0.5cm}
\end{figure}
For the downstream model, we use a latent-conditioned direct regressor. The initial rough channel estimation is first embedded by a $1\times 1$ convolution into a 256-dimensional feature map on the $32\times 16$ grid. In parallel, the predicted latent prior is projected to the same hidden dimension and used as the conditioning signal. The reconstruction network then applies 6 transformer-style blocks, each consisting of self-attention, cross-attention to the latent prior, and a feed-forward layer, followed by a linear projection to the output recovery.

For each subband, pilot positions are sampled with a stride of 8 and a random starting offset. Additive Gaussian noise is applied only to the observed pilot entries, with the noise power sampled uniformly in NMSE range $[0,1]$.

All models are trained with batch size 32, weight decay $10^{-4}$. The learning rate increases linearly from 0 to $10^{-4}$ in the first 10\% of training, and gradually falls to 0 with a cosine scheduler during the following training. We use $\mu=0.95$ for the exponential moving average between teacher and student models during JEPA training.
\subsection{Baseline Methods}
To better evaluate the role of predictive latent modeling, we consider the four baseline methods below.
\begin{itemize}
    \item \textbf{Simple Prediction: }Here, we consider a simple prediction network. We directly use the past channel matrices to recover the next-step channels. It is a non-JEPA baseline that only performs prediction without current estimation.
    \item \textbf{Simple Estimation: }It is another baseline that only uses the current pilot-based sparse and noisy estimation to recover the target channel without past information.
    \item \textbf{Direct Training: }It is an ablation method for the JEPA double-stage train scheme. Instead, we directly train the overall network with the final reconstruction target for 400 epochs.
    \item \textbf{Current-Latent Reconstruction: }It is an ideal upper-bound method which is unrealistic in practice. Instead of using predicted latents, we directly map the current channel to the latent domain and input it into the same downstream network. This is to show the upper bound of the investigated scheme with perfect prediction.
\end{itemize}
\subsection{Effect of Different Regularization Items in Pretraining}
Fig.~\ref{fig:jepa_results} shows the pretraining results under different regularization items with the same network structure. Although the JEPA objective is optimized with an L-1 latent prediction loss, we report NMSE in the latent prediction evaluation because it is more directly aligned with downstream wireless reconstruction quality. Here $\lambda_\textrm{g}$ takes value of 0.2 when enabled.

A moderate covariance regularization is clearly necessary, and the best result is obtained with $\lambda_{\mathrm{cov}}=0.1$. In contrast, among the three recovery-oriented regularizations, the performance degrades from diffusion to amplitude recovery and then to direct recovery, indicating that finer detail restoration makes the JEPA prediction task harder.

The three grounding variants impose different levels of CSI-detail preservation. Direct CSI recovery is the strongest constraint because it encourages the latent code to preserve full complex channel coefficients. Amplitude recovery relaxes the phase requirement but still constrains channel-strength details. Diffusion grounding is weaker because the denoising objective does not require the latent code to deterministically reconstruct all CSI coefficients. However, among the grounding variants, diffusion grounding is the least harmful because the denoising target can be partially solved from the corrupted sample itself, so the latent condition is not forced to encode all fine-grained CSI details.

This is also consistent with our dataset, since phase-level prediction is inherently difficult due to the position shift. More importantly, latent prediction quality should be interpreted together with downstream reconstruction. Poor JEPA prediction NMSE indicates weak predictive representations, whereas good JEPA prediction NMSE is only a necessary, not sufficient, condition for downstream gains.

\subsection{Channel Prediction and Reconstruction Results}
\begin{figure}[t]
    \centering
    \includegraphics[width=0.9\linewidth]{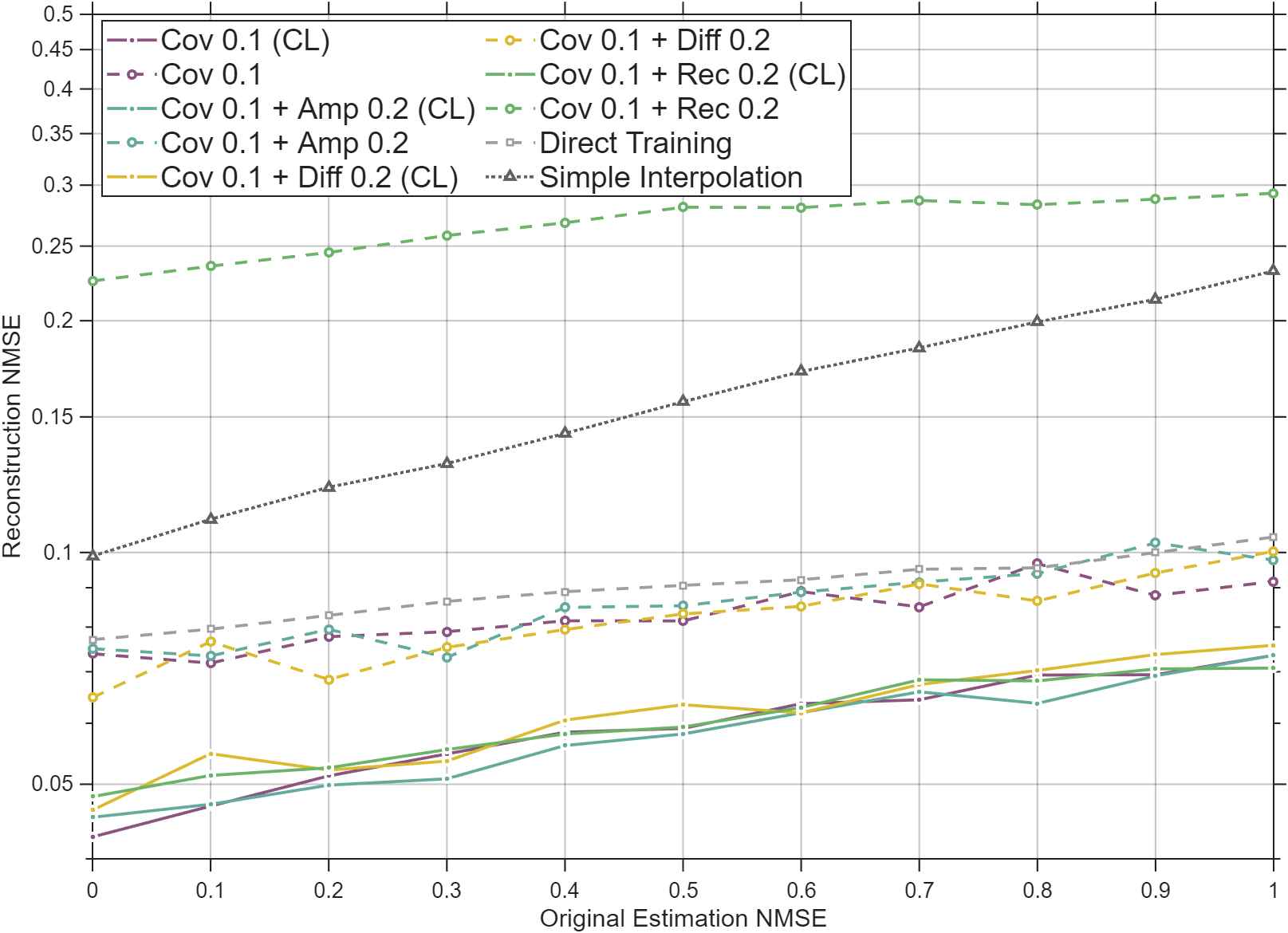}
    \caption{Downstream channel reconstruction performances of different methods. We note that the simple prediction experiment fails to converge.}
    \label{fig:downstream_recon_nmse_8curves}
    \vspace{-0.5cm}
\end{figure}
Fig.~\ref{fig:downstream_recon_nmse_8curves} compares the downstream reconstruction NMSE under different noise levels for different JEPA and baseline methods. We first find that the current-latent reconstruction method are exceptionally good while close to each other. Thus, the quality of learned latents are similar, and the difference mainly lies in the JEPA prediction accuracy. Among the prediction-aided results, using only ``cov'' and ``cov + diff'' yield the best reconstruction performance across the full noise range, which performs similarly. In contrast, adding amplitude recovery or direct recovery consistently increases the reconstruction NMSE. This indicates that recovery-oriented auxiliary regularization does not improve the downstream usefulness of the learned JEPA representation, and even sometimes harms the pilot-based reconstruction method. Instead, the most effective choice is a predictive objective regularized only by a moderate covariance term without any recovery-aimed regularization due to the good JEPA prediction results..

Meanwhile, the simple prediction baseline fails to converge due to the random disturbance in the locations. Moreover, the comparisons to the baselines shows that the best JEPA-based method clearly outperforms both simple estimation and direct training baselines, demonstrating the usefulness of the integrated prediction and estimation scheme and the training method.
\section{Conclusion}

In this paper, we proposed a JEPA-inspired FWM for sparse-pilot MIMO-OFDM channel reconstruction. Rather than directly predicting raw MIMO-OFDM channel coefficients, the framework maps time-frequency CSI observations into a compact latent field and performs prediction in the latent space. This representation emphasizes the dominant and relatively stable and predictable propagation-field structure, while reducing sensitivity to unstable small-scale fading and phase-related details. The predicted latent state is then combined with a small number of real-time pilots as a structured prior for channel reconstruction, improving channel reconstruction under sparse pilot observations. The ablation results further show that CSI-domain grounding objectives, including diffusion grounding, amplitude recovery, and direct CSI recovery, do not necessarily improve downstream reconstruction. Instead, forcing the latent representation to preserve excessive fine-grained CSI details can reduce its predictability and practical usefulness as a reconstruction prior.

%Future work may extend this framework toward unified wireless world models for heterogeneous systems and tasks. Modality-specific tokenizer heads can be designed to map CSI observations with different subcarrier spacings, antenna configurations, array geometries, and data formats into a shared latent space. Moreover, task-specific heads may further support resource allocation, beamforming, sensing, localization, and other wireless tasks that rely primarily on large-scale channel information, leading to a representation-centric framework for intelligent wireless communication systems.

\printbibliography
\end{document}